\documentclass[preprintnumbers,amsmath,amssymb,12pt]{revtex4}
\usepackage{bm}%

\begin{document}
\title{$\hbar$-independent Universality of the Quantum-Classical Canonical
Transformations}

\author{T. Hakio\u{g}lu}

\address{(\it{Corresponding author})\\
Physics Department,Bilkent
University, 06533 Ankara, Turkey \\
hakioglu@fen.bilkent.edu.tr}

\author{A. Te\u{g}men}
\address{Department of Physics, Ankara University, Faculty of Sciences\\
06100, Tando\u{g}an-Ankara, Turkey \\
tegmen@science.ankara.edu.tr}%

\author{B. Demircio\u{g}lu}
\address{Sarayk\"{o}y Nuclear Research and Training Center\\
06983, Kazan, Ankara, Turkey \\
bengu@taek.gov.tr}%

\date{\today}
\begin{abstract}
A theory of non-unitary-invertible as well as unitary canonical
transformations is formulated in the context of Weyl's phase space
representations. That all quantum canonical transformations
without an explicit $\hbar$ dependence are also classical
mechanical and vice versa is demonstrated in the phase space.
Contrary to some earlier results, it is also shown that the
quantum generators and their classical counterparts are identical
and $\hbar$-independent. The latter is a powerful result bringing
the theory of classical canonical transformations and the
$\hbar$-independent quantum ones on an equal footing.
\end{abstract}
\maketitle

PACS numbers: 03.65.-w; 02.30.Uu; 04.60 Ds,
\section{Introduction}
Canonical transformations (CTs) played a crucial role in the
 historical development of
quantum mechanics.\cite{e1,e2} So profound the contribution of the
transformation theory to the fundamental understanding  of quantum
mechanics is that it is just to compare it\cite{e3} to the
beginning of a new phase in analytical dynamics initiated by
Poisson in the generalized coordinates and later by Jacobi,
Poincar\'e, Appell and Hamilton in the development of the
canonical formalism. While the development in the early phases of
quantum mechanics was characterized by the configuration and phase
space approaches, its later elaborations led to the conception of
abstract Hilbert space through which the formerly important
transformation theory approach lost its {\it momentum}\cite{e3}.
Contrary to the case with the well-formulated linear
CTs\cite{KBW}, formulating the nonlinear ones is made more
challenging in the presence of deep problems as invertibility,
uniqueness\cite{e1}, unitarity versus non-unitarity\cite{e1,Mosh},
 and, in many cases,
even the lack of the transformation generators in connection
with the absence of the identity limit\cite{Anderson1}. They
mediate a unique language with the path integral quantization at
one extreme\cite{Barut,Hietarinta} and the Fresnel's geometrical
optics on the other\cite{Guillemin}. Their unitary representations
were first treated by Dirac\cite{e2} as a first step towards the
path integral quantization.

In 1927 Weyl\cite{Weyl} introduced a new quantization scheme
based on a generalized operator Fourier correspondence between an
operator $\hat{\cal F}={\cal F}(\hat{p},\hat{q})$ and a phase
space function $f(p,q)$. To observe the Dirac correspondence as a
special case, Weyl restricted the space of the operator to the
Hilbert-Schmidt space where monomials such as $\hat{p}^m \,
\hat{q}^n$ acquire finite norm for all $0 \le m,n$. Weyl's
formalism was then extended by the independent works of von
Neumann, Wigner, Groenewold and Moyal \cite{vNWGM} to a general
phase space correspondence principle between the operator formulation
of quantum mechanics and its equivalent version on the non-commutative
phase space.

There has been some reviving interest in the quantum CTs and their
classical limits\cite{Anderson1,Kyoto,CZ2,Ghandour}. The goal of
this paper is to formulate the quantum CTs within phase space
covariant formulation of Weyl quantization. More importantly, it
is also shown that the Weyl quantization allows (contrary to some
conventional belief, see Ref.\,[6]) a restricted covariance under
certain types of nonlinear CTs.

\section{Weyl quantization and canonical transforms} According to
the Weyl scheme a Hilbert-Schmidt operator $\hat{\cal F}$ is
mapped one-to-one and onto to a phase space function $f(p,q)$ as
\begin{subequations}\label{corr.1}
\begin{equation}
f(p,q)=Tr\Bigl\{\hat{\Delta}(p,q)\,\hat{\cal F}\Bigr\},
\end{equation}
\begin{equation}
\hat{\cal F}=\int\,\frac{1}{\hbar}\frac{dp}{2\pi} \frac{dq}{2\pi}\,f(p,q)\,
\hat{\Delta}(p,q)\;,\qquad -\infty\;<\,p,q\,<\;\infty
\end{equation}
\end{subequations}
where
\begin{equation}
\hat{\Delta}(p,q)=\int\,d\alpha d\beta\,e^{-i(\alpha p+\beta
q)/\hbar}\, e^{i(\alpha \hat{p}+\beta \hat{q})/\hbar}\,,\qquad
-\infty\;<\,\alpha,\beta\,<\;\infty \label{corr.2}
\end{equation}
is an operator basis satisfying all the necessary conditions of
completeness and orthogonality of the generalized Fourier operator
expansion. The phase space function $f(p,q)$ is often referred to
as the phase space symbol of $\hat{\cal F}$. The operator product
corresponds to the non-commutative, associative $\star$product
\begin{eqnarray}
\hat{\cal F}\,\hat{\cal G} \quad \Longleftrightarrow  \quad
f~\star ~ g ~, \qquad
\hat{\cal F} \, \hat{\cal G} \, \hat{\cal H} \quad
\Longleftrightarrow  \quad f~\star ~ g ~\star ~h
\label{corr.3a}
\end{eqnarray}
where $\hat{\cal F}~,\hat{\cal G}~,\hat{\cal H}$ and their
respective symbols $f,~g,~h$ are defined by (\ref{corr.1}) and
(\ref{corr.2}). The {\it $\star$-product} is a formal exponentiation
of the Poisson bracket $\stackrel{\leftrightarrow}{\cal
D}_{(q,p)}$ as
\begin{equation}
\star_{(q,p)} \equiv exp\Bigl\{{i\hbar \over 2}\,
\stackrel{\leftrightarrow}{\cal
D}_{(q,p)}\Bigr\}=\sum_{n=0}^{\infty}\, ({i\hbar \over 2})^n\,{1
\over n!}\, \Bigl[\stackrel{\leftrightarrow}{\cal
D}_{(q,p)}\Bigr]^n, \qquad
\stackrel{\leftrightarrow}{\cal D}_{(q,p)}=
{\stackrel{\gets}{\partial} \over \partial q}\,
{\stackrel{\to}{\partial} \over \partial p}-
{\stackrel{\gets}{\partial} \over \partial p}\,
{\stackrel{\to}{\partial} \over \partial q}
\label{star.1}
\end{equation}
where the arrows indicate the direction that the partial
derivatives act. Unless specified by arrows as in (\ref{star.1}),
their action is implied to be on the functions on their right.
According to (\ref{corr.3a}) the symbol of the
commutator is defined by the Moyal bracket
$$
[\hat{\cal
F},\hat{\cal G}] \quad \Leftrightarrow  \quad
\{f(p,q),g(p,q)\}^{(M)}_{q,p}=f \star_{q,p} g- g \star_{q,p} f
$$
which has a crucial role in deformation
quantization.\cite{defquant} In the latter, the Moyal bracket is a
representation of the quantum commutator in terms of a nonlinear
partial differential operator, and at the same time it is an
$\hbar$-deformation of the classical Poisson bracket. The
canonical commutation relation (CCR) between the canonical
operators, say $\hat{P}, \hat{Q}$, is represented by the phase
space symbols of these operators denoted respectively by $P(p,q),
Q(p,q)$. If $[\hat{P},\hat{Q}]=-i\hbar$ then
\begin{eqnarray}
& &\{P,Q\}^{(M)}_{q,p} \nonumber \\
&=&2\,\sum_{k=0}^{\infty}\,({i\hbar \over
2})^{2k+1}\, {1 \over
(2k+1)!}\,P(p,q)\,\Bigl[\stackrel{\leftrightarrow}{\cal D}_{(q,p)}
\Bigr]^{2k+1}\,Q(p,q) \nonumber \\
&=&-i\hbar~. \label{mb.2}
\end{eqnarray}
It is well known that, a large class of CT can be represented by
not only unitary but also non-unitary (and invertible)
operators\cite{e1} whose action preserve the CCR. Counter examples
to unitary transformations\cite{Mosh} are abound and some of the
distinct ones are connected with the multi-valued (non-invertible)
or domain non-preserving (non-unitary and invertible) operators. A
few examples can be given by the polar-phase-space\cite{Mosh}
(i.e. action-angle) and quantum Liouville transformation\cite{CZ}
which are multi-valued transformations, or Darboux type
transformations between iso-spectral Hamiltonians\cite{Anderson1}.

Here we will reformulate the quantum canonical (unitary
as well as non-unitary) transformations within the Weyl formalism
paying specific attention to a particular subclass of them
characterized by no explicit $\hbar$ dependence in the
canonical variables $P(p,q)$ and $Q(p,q)$.
The importance of this particular class is that, thinking
of $\hbar$ as a free parameter, the only
non-zero contribution to the $\hbar$ expansion of the {\it
canonical} Moyal bracket in (\ref{mb.2}) is the first (i.e. $k=0$)
term
\begin{eqnarray}
\{P,Q\}^{(M)}_{q,p}=i\hbar\,\{P,Q\}^{(P)}_{q,p}+{\cal
O}(\hbar^{2k+1}) \Bigl\vert_{1 \le k}  \mapsto  -i\hbar
\label{central0}
\end{eqnarray}
yielding
\begin{equation}
\{P,Q\}^{(M)}_{q,p} =i\,\hbar\,\{P,Q\}^{(P)}_{q,p}
\label{central1}
\end{equation}
where all ${\cal O}(\hbar^{2k+1})$ terms with $1 \le k$
necessarily vanish. (In Eq's\,(\ref{central0}) and
(\ref{central1}) the superscript $P$ stands for the Poisson
bracket). Eq(\ref{central1}) is the statement that the classical
and quantum canonical $\hbar$-independent transformations are
identical in the group theory sense yielding the strong result
that their generating functions should also be identical. From the
Lie algebraic perspective, the equivalence of the classical and
quantum generators has been established in Ref.\cite{HD}. This
proof obviously contradicts with some earlier
results\cite{Ghandour,Dragt} in which the Moyal covariance stated
in (\ref{central1}) was overlooked.

We also observe that (\ref{central1}) holds between the canonical
pairs, whereas it is not generally true for arbitrary functions
$f(p,q)$ and $g(p,q)$. Eq.\,(\ref{central1}) states an equivalence
between the canonical Moyal and the canonical Poisson brackets for
$\hbar$ independent transformations.

The result in (\ref{central1}) implies that an $\hbar$ independent
quantum CT is also a classical CT, a result that was obtained by
Jordan\cite{e1} long time ago using a semiclassical approach.

\section{The phase space images of canonical transformations} The
Weyl formalism is restricted to a subspace of the Hilbert space in
which the state functions decay sufficiently strongly at the
boundaries to admit an infinite set of finite valued phase space
moments $\hat{p}^m\,\hat{q}^{n}$ with non-negative integers $m,n$.
If the moments are symmetrically ordered (i.e. Weyl ordering)
we denote them by
$\hat{t}_{m,n}=\{\hat{p}^m\,\hat{q}^n\}$. The $\hat{t}_{m,n}$'s
are simpler to represent in the phase space and they correspond to
the monomials $p^m\,q^n$. A function $f(p,q)$ which can be written
as a double Taylor expansion in terms of the monomials $p^m\,q^n$
corresponds to a symmetrically ordered expansion of an operator
$\hat{\cal F}$ as
\begin{equation}
f(p,q)=\sum_{0 \le (m,n)}~f_{m,n}\,p^m\,q^n \quad \Leftrightarrow \quad
\hat{\cal F}=\sum_{0 \le (m,n)}~f_{m,n}\,\hat{t}_{m,n}^{(0)}~.
\label{wcl.2}
\end{equation}
Symmetrically ordered monomials are Hermitian and they can be
convenient in the expansion of other Hermitian operators.

The phase space representations are more convenient to use than
the operator algebra for keeping track of $\hbar$'s. Since
$\hat{t}_{m,n} \Longleftrightarrow p^m\,q^n$, intrinsic $\hbar$
dependencies appear only in the phase space expansions
representing non-symmetrical monomials. Suppose that the operator
$\hat{\cal F}$, which has the Weyl representation $f(p,q)$, is
transformed by an operator $\hat{U}$ which has the Weyl
representation $u(p,q)$ by $\hat{\cal
F}^\prime=\hat{U}^{-1}\,\hat{\cal F}\,\hat{U}$. Assume that the
transformation $\hat{U}$ is given in an exponential form
$\hat{U}_{\cal A}=e^{i\,\gamma \,\hat{\cal A}/\hbar}$ where $\gamma$ is
a continuous parameter and the generator $\hat{\cal A}={\cal
A}(\hat{p},\hat{q})$ is expanded ala (\ref{wcl.2}) as
\begin{equation}
{\cal A}(\hat{p},\hat{q}) =\sum_{m,n}\,a_{m,n}\,\hat{t}_{m\,n}^{(0)}
\label{bopp.1}
\end{equation}
where $a_{m,n}$'s are the expansion coefficients. We than have
by $\hat{\cal F}^\prime=\hat{U}^{-1}\,\hat{\cal F}\,\hat{U}$
and Eq.(\ref{corr.1})
\begin{eqnarray}
f^\prime(p,q)&=&Tr\{\hat{\cal F}^\prime\,\hat{\Delta}\}=Tr\{\hat{\cal F}\,
\hat{U}_{\cal A}\,\hat{\Delta}\,\hat{U}_{\cal A}^{-1}\} \nonumber \\
\hat{U}_{\cal A}\,\hat{\Delta}\,\hat{U}_{\cal
A}^{-1}&=&\hat{\Delta}+ \frac{i\gamma}{\hbar} \, [\hat{\cal
A},\hat{\Delta}]+{(i\gamma)^2 \over 2! \hbar^2}\,[\hat{\cal A}, [\hat{\cal
A},\hat{\Delta}]] +\cdots \label{bopp.2}
\end{eqnarray}
The right hand side of (\ref{bopp.2}) can be represented by
certain linear first order phase space differential operators
producing the left and right action of $\hat{p}$ and $\hat{q}$ on
$\hat{\Delta}$ as\cite{Vercin}
\begin{subequations}\label{corr.6}
\begin{equation}
\hat{p}\,\hat{\Delta}(p,q)=\underbrace{ [p+{i\hbar \over
2}\,{\partial \over \partial q}]}_{\hat{p}_L}\hat{\Delta}(p,q)~,
\qquad \hat{\Delta}(p,q)\,\hat{p}=\underbrace{ [p-{i\hbar \over
2}\,{\partial \over \partial q}]}_{\hat{p}_R}\, \hat{\Delta}(p,q)
\end{equation}
\begin{equation}
\hat{q}\,\hat{\Delta}(p,q)=\underbrace{ [q-{i\hbar \over
}\,{\partial \over \partial p}]}_{\hat{q}_L}\, \hat{\Delta}(p,q)~,\qquad
\hat{\Delta}(p,q)\,\hat{q}=\underbrace{ [q+{i\hbar \over
2}\,{\partial \over \partial p}]}_{\hat{q}_R}\, \hat{\Delta}(p,q)
\end{equation}
\end{subequations}
and thus,
\begin{eqnarray}\label{corr.7}
\bigl[\hat{t}_{m,n},\hat{\Delta}(p,q)]&=&\Big\{\hat{p}_L^m\,\hat{q}_L^n-
\hat{p}_R^m\,\hat{q}_R^n\Bigr\}\, \hat{\Delta}(p,q) \nonumber \\
&\equiv & \hat{S}_{m,n}\,\hat{\Delta}(p,q)
\end{eqnarray}
where we used the specific notation $\hat{S}_{m,n}$ for the image
of the symmetric monomials $\hat{t}_{m,n}$. Using
Eqs.\,(\ref{corr.6}), the first commutator in the expansion in
(\ref{bopp.2}) becomes
\begin{equation}
[\hat{\cal A},\hat{\Delta}]=\hat{V}_{\cal A}\,\hat{\Delta}(p,q)
\label{algebra.1}
\end{equation}
where $\hat{V}_{\cal A}$ is the Moyal-Lie representation\cite{HD} of
the generator ${\cal A}$ given by
\begin{equation}
\hat{V}_{\cal A}=\sum_{m,n}\,a_{m,n}\,
\Bigl\{\hat{p}_{L}^m \hat{q}_{L}^n - \hat{p}_{R}^m\hat{q}_{R}^n \Bigr\}
\label{mlt.1}
\end{equation}

The right hand side of (\ref{bopp.2}) can be obtained by
infinitely iterating the commutator (\ref{algebra.1}) which yields
\begin{equation}
\hat{U}_{\cal A}\,\hat{\Delta}\,\hat{U}_{\cal A}^{-1}=
e^{i\gamma\,\hat{V}_{\cal A}/\hbar}\,\hat{\Delta}~. \label{bopp.5}
\end{equation}
Using Eq.\,(\ref{bopp.5}) in (\ref{bopp.2})
\begin{equation}
f^\prime(p,q)=e^{i\gamma \,\hat{V}_{\cal A}/\hbar}\,f(p,q)~.
\label{bopp.6}
\end{equation}
There exists a linear map, for given
$\hat{\cal A}$, such that $[\quad ,\hat{\Delta}] :\hat{\cal
A}\mapsto \hat{V}_{\cal A}\hat{\Delta}$. It is trivial that
$\hat{\cal C}=\alpha\,\hat{\cal A}+\beta\, \hat{\cal B}$ is mapped
as $\hat{V}_{\cal C}=\alpha\,\hat{V}_{\cal A}+
\beta\,\hat{V}_{\cal B}$. Thus $[\hat{\cal A},\hat{\cal B}]$ is
mapped as
\begin{equation}
\hat{V}_{[{\cal A},{\cal B}]}= -[\hat{V}_{\cal A},\hat{V}_{\cal
B}] \label{bopp.6b}
\end{equation}
via the Jacobi identity. Hence, if the closed set
$\{\hat{\cal A}_{i}\}$ are
generators of a Lie algebra then their images $\hat{V}_{{\cal
A}_i}$ are generators of the Moyal-Lie algebra\cite{HD}.

The Weyl correspondence including the covariance under canonical
transformations can now be summarized in the commuting diagram
\begin{equation}
\begin{array}{rlrlrlrl}
& f(p,q) ~~~~~~ & \stackrel{Weyl}{\Longleftrightarrow}& ~~~~~~~
                                                    & \hat{\cal F}
\\
\hat{V}_{\cal A}   &\Updownarrow  & ~~&    & \hat{U}_{\cal A} ~~
\Updownarrow \\
f^\prime&=e^{i\gamma \hat{V}_{\cal A}/\hbar}\,f    &
\stackrel{Weyl}{\Longleftrightarrow}  & ~~~~~~~~~ &\hat{\cal
F}^\prime~.
\end{array}
\label{corr.9}
\end{equation}
The meaning of the diagram (\ref{corr.9}) can be facilitated by an
example. Consider, for instance, the unitary transformation
corresponding to $\hat{U}_{2,1}$. Using Eq.\,(\ref{corr.6}) and
(\ref{corr.7}) we find the corresponding differential generator
$\hat{S}_{2,1}$ as
\begin{equation}
\hat{V}_{\cal A}=\hat{S}_{2,1}=i\hbar\,(2p\,q\,\partial_q -p^2\,\partial_p
+\frac{\hbar^2}{4}\;
\partial_{q}^2\;\partial_p) \label{corr.10}
\end{equation}
which has an explicit overall $\hbar$ dependence. Also note that
$\hat{S}_{2,1}$ is an Hamiltonian vector field. For any $f(p,q)$
its action gives the Poisson (and Moyal) bracket
\begin{eqnarray}
\hat{S}_{2,1}\,f(p,q)=i\hbar\{f(p,q),p^2\,q\}^{(P)}=
\{f(p,q),p^2q\}^{(M)}_{q,p}.
\end{eqnarray}
Let us consider for $f$ and $f^\prime$ in the diagram
(\ref{corr.9}) the canonical coordinates $(p,q)$ and $(P,Q)$.
Then, using Eq.\,(\ref{corr.10})
\begin{subequations}\label{corr.11}
\begin{equation}
P(p,q)=e^{-i\gamma \hat{S}_{2,1}/\hbar}\,p={p \over 1+\gamma\, p}
\end{equation}
\begin{equation}
Q(p,q)=e^{-i\gamma \hat{S}_{2,1}/\hbar}\,q=q\,(1+\gamma\,p)^2~,
\end{equation}
\end{subequations}
such that $P^2Q=p^2q$. It can be directly observed that the
canonical transformation in Eq.\,(\ref{corr.11}) respects
(\ref{central1}).

\section{Generating functions}\label{GF}
The Weyl symbol of an admissible operator $\hat{U}$ is given by,
\begin{equation}
\hat{U}=\int\,\frac{dp\,dq}{(2\pi)^2\hbar}\,
u(p,q)\,\hat{\Delta}(p,q)~. \label{int.1}
\end{equation}
Since $\hat{U}$ is unitary, then $u(p,q)$ satisfies
$u^{*}(p,q)=u^{(-1)}(p,q)$ where $*$ denotes the complex
conjugation and the $u^{(-1)}$ is the Weyl symbol of
$\hat{U}^{-1}$. Eq.(\ref{int.1}) also converts an inner
product in the Hilbert space to that in the phase space.
The former is given by
\begin{equation}
(\psi,\hat{U}\,\varphi)=\int\,dq\,\psi^*(q)\,(\hat{U}\,\varphi)(q)~.
=\int\,\frac{dp\,dq}{(2\pi)^2\hbar}\,
u(p,q)\,(\psi,\hat{\Delta}(p,q)\,\varphi)
\label{int.2}
\end{equation}
Using the matrix elements $\langle y\vert \hat{\Delta}(p,q)\vert x
\rangle$ and considering a functional derivative of (\ref{int.2})
with respect to $\psi^*(y)$, we find in the coordinate-coordinate
representation that
\begin{subequations}\label{int.3}
\begin{equation}
(\hat{U}\,\varphi)(y)=\int\,dx \,e^{iF(y,x)}\,\varphi(x),
\end{equation}
\begin{equation}
e^{iF(y,x)}=\int\,{dp \over 2\pi \hbar}\,e^{-ip\,(x-y)/\hbar}\,
u(p,{x+y \over 2}).
\end{equation}
\end{subequations}
For the mixed (coordinate-momentum) representation
\begin{subequations}\label{int.4}
\begin{equation}
(\hat{U}\,\varphi)(y)=\int\,{dp_x \over 2\pi \hbar}\,
e^{i\,K(y,p_x)}\,\tilde{\varphi}(p_x),
\end{equation}
\begin{equation}
e^{i\,K(y,p_x)}=\int\,dx \,e^{i[F(y,x)+x\,p_x/\hbar]},
\end{equation}
\end{subequations}
alternatively, in the momentum-momentum representation we have
\begin{subequations}\label{int.5}
\begin{equation}
({\hat{U}\,\tilde{\varphi}})(p_y)=\int\,\frac{dp_x}{2\pi
\hbar}\,e^{i\,H(p_y,p_x)} \,\tilde{\varphi}(p_x),
\end{equation}
\begin{equation}
e^{i\,H(p_y,p_x)}=\int\,dq \, e^{-i\,q(p_x-p_y)/\hbar}\,
u({p_y+p_x \over 2},q).
\end{equation}
\end{subequations}
For the other mixed case
\begin{subequations}\label{int.6}
\begin{equation}
({\hat{U}\,\tilde{\varphi}})(p_y)=\int\;
dx\;e^{i\,L(p_y,x)}\,\varphi(x),
\end{equation}
\begin{equation}
e^{i\,L(p_y,x)}=\int
\frac{dp_x}{2\pi\hbar}\,e^{i\,[H(p_y,p_x)-x\,p_x/\hbar]}.
\end{equation}
\end{subequations}
Hilbert space representations of canonical transformations like
(\ref{int.3})-(\ref{int.6}) have been written by Dirac using intuitive
arguments in his celebrated book on quantum
mechanics\cite{e2}. Here a direct proof of his results is presented using
the Weyl correspondence.

Note, that we have not assumed any particular property for the
generic unitary operator $\hat{U}$. Now we assume that $\hat{U}$
produces the canonical transformation
\begin{equation}
\hat{P}=\hat{U}^{-1}\,\hat{p}\,\hat{U}~,\qquad
\hat{Q}=\hat{U}^{-1}\,\hat{q}\,\hat{U}. \label{int.7}
\end{equation}
Multiplying both sides by $\hat{U}$ on the left and using the Weyl
correspondence in Eq.\,(\ref{corr.3a}) we find
\begin{subequations}
\begin{eqnarray}
u(p,q) \star Q(p,q)&=& q \star u(p,q) = \Bigl(\, q+{i\hbar \over
2}\,{\partial \over \partial p}\Bigr)\,u~,
\label{int.7a} \\
u(p,q) \star P(p,q) &=& p \star u(p,q) = \Bigl(\, p-{i\hbar \over
2}\,{\partial \over \partial q}\Bigr)\,u\,, \label{int.7b}
\end{eqnarray}
\end{subequations}
where $\star=\star_{q,p}$ as defined in (\ref{star.1}). Another
crucial property of the $\star$-product is that, $\star=\star_{q,p}=\star_{Q,P}$. This can be easily seen from (\ref{star.1}) considering
that $p,q$ and $P,Q$ are related by a CT.
Once Eq's(\ref{int.7b}) are solved, the generators of the CT
can be found by using Eq's.(\ref{int.3})-(\ref{int.6}).

\section{Examples}

Let us solve the Eq's\,(\ref{int.7a}) and (\ref{int.7b}) for a few
well known cases. We first do it for the group of linear
symplectic transformations $SL_2(\mathbb{R})$.

a) $SL_2(\mathbb{R})$:

In this case we have
\begin{eqnarray}\label{int.8}
\left(
\begin{array}{c}
P \\ Q
\end{array}\right)=g
\left(
\begin{array}{c}
p \\ q
\end{array}\right)\;,\quad
g= \left(
\begin{array}{cc}
a & b \\ c & d
\end{array}\right)
\in SL_2(\mathbb{R}).
\end{eqnarray}
Directly using (\ref{int.8}) in (\ref{int.7a}) and (\ref{int.7b})
one has
\begin{equation}\label{int.9}
u(p,q)={2 \over \sqrt{a+d+2}}
 \exp\Bigl\{{-2\,i \over
(a+d+2)\,\hbar}\,[b\,q^2+c\,p^2-(a-d)\,p\,q] \Bigr\}
\end{equation}
where $Tr{g} \ne -2$ and the normalization is chosen such that
identity transformation is $u(p,q)=1$. By (\ref{int.3})
this can be converted into the kernel
\begin{equation}
e^{i\,F(y,x)}={e^{-i\pi/4} \over \sqrt{2\pi\hbar\,c}}\, e^{{-i
\over 2\hbar \, c}(ay^2+dx^2-2xy)} \label{int.10}
\end{equation}
yielding the correct integral kernel for $SL_2(\mathbb{R})$
transformation including the normalization factor\cite{KBW}. The
special cases such as $Tr{g}=-2$ can be treated with additional
limiting procedures which will not be considered here.

b) Linear Potential:

The second exactly solvable system is the linear potential model
\begin{equation}
{P \choose Q}={p \choose q+ap^2}~,\qquad a \in
\mathbb{R}\label{int.11}
\end{equation}
using (\ref{int.7a}) and (\ref{int.7b}) once more we find,
\begin{equation}
u(p,q)=N_a \, exp(-{i\,a \over 3 \hbar}\,p^3)~,\qquad
N_a\Bigr\vert_{a=0}=1 \label{int.12}
\end{equation}
which is more conveniently used in a mixed type of transformation
kernel given by Eq.\,(\ref{int.4}) as
\begin{equation}
e^{i\,K(y,p_x)}=e^{{-i \over \hbar}\,(y\,p_x-{a \over 3}\,p_x^3)}
\label{int.13}
\end{equation}
where $N_a=1$ is used, yielding the correct solution of the linear
potential model.\cite{CZ} Also unphysical $\hbar$ dependencies may
appear if the Moyal covariance is not correctly taken into
account\cite{Dragt}.

In both examples the unitary transformation kernel $u(p,q)$ is
closely related to the appropriate classical generating function
of the canonical transform as remarked by Dirac\cite{e2} in the
early days of the quantum theory. A close look into (\ref{int.10})
as well as (\ref{int.13}) confirms that they are exponentiated
versions of one of the four types of generating functions that one
learns in the textbooks. An important remark is that, since the
quantum and classical generating functions are identical, there
are no $\hbar$-corrections as anticipated in some earlier
works\cite{Dragt}. Indeed, (\ref{int.10}) is, after renaming $y
\to Q$ and $x \to q$ as the new and the old coordinates
\begin{equation}
F_1^{(q)}(Q,q)=-{1 \over 2c}\,(a\,Q^2+d\,q^2-2\,Q\,q)
\label{f1}
\end{equation}
which is just the classical generating function $F_{1}^{(cl)}(Q,q)$ for
the linear symplectic transformations satisfying $p=\partial
F_{1}^{(cl)}(Q,q)/\partial q$ and $P=-\partial F_{1}^{(cl)}(Q,q)/\partial Q$.

Likewise, in Eq(\ref{int.13}) the quantum generator is (in the notation
$y \to Q$ and $p_x \to p$,
\begin{equation}
F_3^{(q)}(Q,p)=-Q\,p+\frac{a}{3}p^3
\label{f3}
\end{equation}
which is just the classical generating function $F_3^{(cl)}(Q,p)$
for the nonlinear transformation in Eq.(\ref{int.11}) satisfying
$q=-\partial F_3^{(cl)}(Q,p)/\partial p$ and
$P=-\partial F_3^{(cl)}(Q,p)/\partial Q$. Eq.\,(\ref{int.13}) that
was found for the linear potential model matches exactly with the
exponentiated classical generator and agrees with Dirac's
exponentiation formula\cite{e2}.

%
Eq.\,(\ref{central1}) provides some background we need in order to
understand the solutions of (\ref{int.7a}) and (\ref{int.7b}) for
the class of problems for which $u(p,q)$ has no $\hbar-{\it
corrections}$. The $\hbar$-corrections to the CT generators were
proposed in Ref.[15] in reference to a particular Hamiltonian.
This concept can be made independent of a dynamical model by
demanding that the solution of (\ref{int.7a}) and (\ref{int.7b})
yields integral kernels $F_1(Q,q),F_2(q,P),F_3(Q,p),F_4(P,p)$
in (\ref{int.3})-(\ref{int.6}) which are all in the order of
$1/\hbar$ independent from any class of Hamiltonians considered
implied by
\begin{equation}
u(p,q)=e^{{2\,i \over \hbar}\,T(p,q)}~,\qquad {\partial T \over
\partial \hbar}=0
\label{int.15b}
\end{equation}
hence $T(p,q)$ has no $\hbar$ dependence and the corresponding
generating functions $F_1,F_2,F_3,F_4$ in (\ref{int.3})-(\ref{int.6})
are identical to their classical counterparts.

By inspecting Eq\,s\,(\ref{int.7a}) and (\ref{int.7b}) one expects
to find that the particular class of transformations for which
\begin{eqnarray}
u(p,q) \star_{q,p} Q(p,q)&=&u(p,q) \star_{Q,P} Q \label{int.15.c.a} \\
u(p,q) \star_{q,p} P(p,q)&=&u(p,q) \star_{Q,P} P
\label{int.15.c.b}
\end{eqnarray}
holds, yields $\hbar-{\it uncorrected}$ solutions as in
Eq.\,(\ref{int.15b}) for $u(p,q)$. It is intuitive that the
conditions in (\ref{int.15.c.a}) and (\ref{int.15.c.b}) are
sufficient but not necessary for the $\hbar$-uncorrected solutions
in (\ref{int.15b}). If Eq's\,(\ref{int.15.c.a}) and
(\ref{int.15.c.b}) hold, then
\begin{eqnarray}
(Q-{i\hbar \over 2}\partial_{P})\,u(p,q)&=&(q+{i\hbar \over 2}\,
\partial_{p})\,u(p,q) \label{int.16a} \\
(P+{i\hbar \over 2}\partial_{Q})\,u(p,q)&=&(p-{i\hbar \over 2}\,
\partial_{q})\,u(p,q)~.
\label{int.16b}
\end{eqnarray}
Considering the general form in (\ref{int.15b}) the solution is
\begin{eqnarray}\label{int.18}
\left(
\begin{array}{c}
\partial_p\\ \partial_q
\end{array}\right)T &=&
(2+\partial_P p+\partial_Q q)^{-1} \nonumber\\
&\times &\left(
\begin{array}{cc}
1+\partial_Q q & -\partial_P q \\
-\partial_Q p & 1+\partial_P p
\end{array}\right)
\left(
\begin{array}{c}
q-Q\\P-p
\end{array}\right)
\end{eqnarray}
here it is required that the determinant of the matrix
$(2+\partial_{P}\,p+\partial_{Q}\,q)$ is non-zero and we employed
the Lagrange bracket $\{q,p\}_{Q,P}=1$ as a canonical invariant.
The solution to (\ref{int.18}) is clearly $\hbar$ independent if
the canonical transformation $(p,q) ~~\mapsto ~~(P,Q)$ is also
independent of $\hbar$. Eq's\,(\ref{int.16a}) and (\ref{int.16b})
are manifestly satisfied for the linear symplectic transformations
in Eq.\,(\ref{int.8}).

\section{Conclusions}
In this work we introduced Weyl's phase space representations of
the nonlinear quantum canonical transformations. We have shown that
the nonlinear canonical transformations which generally lack unitary
representations in Hilbert space, have unitary phase space
representations.

It has been believed for a long time that Weyl quantization did
not possess covariance under nonlinear CT. As the results in this
work indicate, different Weyl representations can be connected by
the nonlinear CT thereby extending the concept of covariance
instead of breaking it. Another advantage in seeing this as an
extended covariance is that the presented approach also unifies
with Dirac's transformation theory which is essentially a Hilbert
space approach. Dirac's transformation theory can be naturally
merged [as shown in section (\ref{GF})] with Weyl's phase space
approach bringing the theory of CT (particularly nonlinear,
invertible) back to where it should belong.

Nearly as old as the quantum mechanics itself, the Weyl
quantization remains to be one of the most active fields in a wide
area of physics. Without need of mentioning its applications in
quantum and classical optics, condensed matter physics and
engineering\cite{feature}, it has been put into a more general
frame in the deformation quantization.\cite{defquant} Recently, it
also proved to be an essential part of the non-commutative quantum
field and string theories in the presence of background gauge
fields.\cite{Connes} It is then natural to expect that the theory
of canonical transformations, which is subject to progress within
itself, may also find some applications in these new directions.

\section*{Acknowledgements}
The author T. H. is thankful to C. Zachos (High
Energy Physics Division, Argonne National Laboratory) for
stimulating discussions. This work was supported in part by
T\"{U}B\.{I}TAK (Scientific and Technical Research Council of
Turkey), Bilkent University and
 the U.S. Department of Energy, Division of
High Energy Physics, under contract W-31-109-Eng-38.

\end{document}